\documentstyle[prl,twocolumn,epsfig,aps]{revtex}
 
\newcommand{\be}{\begin{equation}}
\newcommand{\ee}{\end{equation}}
\newcommand{\bea}{\begin{eqnarray}}
\newcommand{\eea}{\end{eqnarray}}

\newcommand{\Dslash}{\mbox{$D$\kern-0.65em \hbox{/}\hspace*{0.25em}}}
\newcommand{\Dslashup}{\not \kern-0.03em D}

\begin{document}
\draft

\title{Efficiency of the UV-filtered Multiboson algorithm}
\author{Constantia Alexandrou$^1$, Philippe de Forcrand$^2$, Massimo D'Elia$^3$
and Haralambos Panagopoulos$^1$}
\address{$^1$Department of Natural Sciences, University of Cyprus, CY-1678 Nicosia, Cyprus}
\address{$^2$ETH-Z\"urich, CH-8092 Z\"urich, Switzerland}
\address{$^3$Dipartimento di Fisica dell'Universit\`a and I.N.F.N., I-56127, Pisa, Italy}

\date{\today}
\maketitle
 
\begin{abstract}
We study the efficiency of an improved Multiboson algorithm 
with two flavours of Wilson fermions in a realistic
physical situation ($\beta = 5.60$, $\kappa = 0.156$
on a $16^3 \times 24$ lattice). 
The performance of this exact algorithm is compared 
with that of a state-of-the-art HMC algorithm:
a considerable improvement is obtained
for the plaquette auto-correlation time,
while the two algorithms
appear similarly efficient at decorrelating
the topological charge.
\end{abstract}
 

\section{Introduction}

The Monte-Carlo simulation of lattice QCD, including the effects of dynamical
quark loops, is a particularly difficult and challenging problem,
as it involves the computation of the fermion
matrix determinant, which appears, after integration over
the anti-commuting fermion fields, in the QCD partition function
\be
{\mathcal Z} = \int \prod_{x,\mu} dU_{x,\mu}~e^{-S_g[U]} ~ 
\prod_{i=1}^{N_f} \det(\Dslash[U] + m_i) \quad.
\ee 
The fermion determinant leads to non-local interactions among the
gauge links $U_{x,\mu}$, so that the cost for updating all links
naively grows at least as ${\mathcal O}(V^2)$, where $V$ is the lattice 
volume.

The standard approach to this problem is given by the Hybrid Monte 
Carlo (HMC) method. The computation of the determinant
is achieved stochastically by the introduction of an auxiliary 
bosonic pseudo-fermion field $\phi$. For the case of two
degenerate flavours one may write
\be
|\det (\Dslash + m)|^2 = \int [d\phi^\dagger] [d\phi]~
e^{- |(\Dslashup + m)^{-1} \phi|^2} \quad.
\label{hmceq}
\ee
One still has to deal with a non-local action involving $(\Dslash + m)^{-1}$,
but now, using Molecular Dynamics, a global updating of $U$ can 
be performed.

An alternative approach, which allows the use of local algorithms,
is the so-called Multiboson method, originally proposed by L\"uscher 
\cite{MB0}. If a polynomial $P_n(x) = c_n \prod_{k=1}^n
(x - z_k)$ can be found which approximates $1/x$ over the whole spectrum of 
$(\Dslash + m)$, then $P_n(\Dslash + m) \approx (\Dslash + m)^{-1}$.
Using a relation similar to Eq. (\ref{hmceq}) one can then write
\bea
\lefteqn{
|\det (\Dslash + m)|^2 \approx |{\det}^{-1} P_n(\Dslash + m)|^2 } \nonumber \\
&~~~~~~~~~~~= c_n^{-V} \int \prod_k [d\phi_k^\dagger] [d\phi_k]~
e^{- \sum_k |(\Dslashup + m - z_k) \phi_k|^2} \quad.
\eea
In this way the original QCD partition function is approximated by 
a functional integration over the gauge links $U$ and $n$ bosonic
fields $\phi_k$, where the integration measure is given in terms
of a local action
\be
S = S_g + \sum_{k=1}^n |(\Dslash + m - z_k) \phi_k|^2 \quad, 
\label{totMBac}
\ee
so that standard powerful local algorithms (heatbath, overrelaxation)
can be used. The systematic error deriving from this approximation
can be corrected either during the simulation, with a global
accept-reject step, or by a reweighting procedure on physical observables.

An important question to the lattice scientific community 
is which of the two approaches (HMC or MB) is more efficient 
for simulating QCD with dynamical fermions.
No definitive answer is yet available. 
On general grounds,
the MB method has the advantage
of being still relatively new, so that
prospects for improvement are much greater. 

The MB approach allows two different strategies for improvement.
The first is the choice of the approximating polynomial.
As the number $n$ of bosonic fields increases, the work per updating
step grows as $n$. Moreover, the autocorrelation time for physical
observables also grows approximately as $n$. It is therefore essential
to choose a polynomial of the lowest possible degree while
preserving sufficient accuracy in the approximation,
i.e. sufficient acceptance in the Metropolis step.
The second is the choice of the update algorithm.
Since we are dealing with a local action, a number of possible efficient local 
algorithms are available. Moreover a global updating of the 
bosonic field variables can be simply implemented, since
their distribution is gaussian.
The coupled dynamics of the system
(gauge links + boson fields) is highly non trivial,
so that the optimal mixture of update algorithms
is essentially determined by numerical experiment \cite{Beat}.
As we will see, a good choice can make a substantial difference.

The aim of the present work is to test the efficiency
of an improved version of the MB algorithm in a physical 
situation and to directly compare its performance with 
a ``state of the art'' HMC algorithm, namely the one used by
the SESAM collaboration \cite{SESAM}.
Clearly, the efficiency of an algorithm is different depending
on the physical observable one is monitoring (i.e., one
obtains different integrated autocorrelation times
for different observables). In the present work
we are studying the plaquette and the topological charge.
These two observables are representative of the smallest- (UV) and 
largest- (IR) scale features of the gauge field. Therefore, their Monte Carlo
dynamics provide a succinct
description of the efficiency of a simulation at decorrelating all
intermediate scales.
The study of the topological charge dynamics is of particular importance: 
because of the associated zero-mode crossing, it is expected theoretically,
and has been shown in practice \cite{HMC1}, that HMC algorithms near
the chiral limit can be particularly inefficient at decorrelating
topology, so that even using a very long simulation time
on supercomputers, one is not able to properly
sample the topological modes of the theory and ensure ergodicity.
The efficiency of the MultiBoson method in this respect has not 
been studied yet.

We have used the exact, non-hermitian version
of the MB algorithm \cite{MB1} with two flavours of Wilson fermions.
In order to reduce the number of bosonic fields we have used the method 
proposed in \cite{PhUV}:
after a preconditioning of the Dirac matrix, which rewrites the effect
of the UV fermionic degrees of freedom as a simple modification of
the pure gauge action (UV-filtering), an optimized polynomial is
constructed numerically by adapting it, via quadratic minimization,
to a typical configuration of the physical ensemble.

The dynamics of the algorithm have then been improved by choosing
a proper mixture of local overrelaxation of the boson and gauge fields
and of global heatbath on the boson fields. The introduction of 
the global heatbath step (whose computational cost is
strongly reduced by using a Quasi-Heatbath with approximate inversion \cite{PhQH})
turns out to be essential in order to improve
the dynamics of the algorithm at small quark mass \cite{borici}.

In Section II we present more details about the algorithm used.
In Section III the numerical results of our simulations are shown.
In Section IV we give our conclusions.


\section{Description of the algorithm}

\subsection{Statics}
We have used the exact, non-hermitian version of the MB algorithm \cite{MB1},
which includes a noisy Metropolis test to correct for the polynomial approximation
to the fermionic determinant \cite{Borici_Ph}.

As the degree $n$ of the approximating polynomial is increased, the work
per independent configuration grows as $n^2$ \cite{MB1}: it is therefore
crucial to keep $n$ as low as possible while maintaining a good approximation,
i.e. a good acceptance for the Metropolis test.

In order to do that we have followed the procedure of \cite{PhUV}. Inspired by
the loop expansion of the fermion determinant,
\be
{\rm det}({\bf 1} - \kappa M) = e^{{\rm Tr~Log}({\bf 1} - \kappa M)}
= e^{- \sum_l \frac{\kappa^l}{l} {\rm Tr}~M^l},
\ee
one rewrites the determinant as
\bea
\lefteqn{ {\rm det}({\bf 1} - \kappa M) = } \\
\label{det}
& e^{- \sum_j a_j {\rm Tr} M^j} \times
{\rm det}\left( ({\bf 1} - \kappa M) ~~e^{+ \sum_j a_j M^j} \right) \nonumber
\eea
where the identity ${\rm det}~e^{A} = e^{{\rm Tr}~A}$ has been used.

The idea behind this is to filter out the UV part of the Dirac spectrum: the term $ e^{- \sum_j a_j {\rm Tr} M^j}$,
which adds a set of small loops to the gauge action, accounts for the UV modes of 
the fermion determinant. In this way the zeros of the new polynomial (i.e. the one
which approximates the inverse of $ ({\bf 1} - \kappa M) ~~e^{+ \sum_j a_j M^j} \>$) can be concentrated near the small eigenvalues
(corresponding to IR modes) and a better approximation can be obtained with less cost.

The number of coefficients $\{a_j\}$ as well as their values can be optimized. 
It is easy to see that for $j \le 4$ the term $e^{-\sum_j a_j {\rm Tr}~M^j }$ can
be reabsorbed in a shift of the coupling $\beta$ of the pure gauge Wilson action,
$\Delta\beta = 192 \kappa^4 a_4$, so that
no computational overhead is incurred.

For the optimization of the coefficients $\{a_j\}$ and of the zeros $\{z_k\}$ of the polynomial
we have followed the procedure suggested in \cite{PhUV}, which we describe here briefly.
The parameters have to be chosen so that $\det W \simeq 1$, where 
$ W \equiv \prod_k^n ({\bf 1} - \kappa M - z_k {\bf 1}) \cdot
({\bf 1} - \kappa M) \cdot e^{\sum_{j=0}^{m-1} a_j M^j} $. A sufficient condition for this is that $W \eta \simeq \eta$ for Gaussian vectors $\eta$.
Therefore one takes an equilibrium gauge configuration, fixes the coefficients $a_j$ to some
initial value, draws one or more Gaussian vectors $\eta$ and
finds, by quadratic minimization, the roots $\{z_k\}$ which minimize the quantity 
$e = \parallel W\eta - \eta \parallel^2 $.

During this procedure also $\partial e/\partial a_j $ can be computed, so that one can repeat this optimization
using new values for $a_j$, and thus minimize also with respect to the $a_j$'s by Newton's method.

In principle one should perform an average over different equilibrium gauge fields. 
In practice results do not change appreciably with the gauge field, so that
one single configuration gives sufficient information.

An interesting byproduct of the optimization is that it is possible to estimate the acceptance
for the Metropolis test expected during the actual
Monte Carlo simulation. The acceptance probability for $\{U_{old}\} \to \{U_{new}\}$ is
\be
min(1,\langle \frac{e^{- |W_{{\rm old}}\>\eta|^2 + |\eta|^2}}
{e^{- |W_{{\rm new}}\>\eta|^2 + |\eta|^2}} \rangle_{\eta})
\ee
and a good estimate for this is ${\rm erfc}(c \parallel W_{{\rm old}}\> \eta - \eta \parallel) $, with $c \sim O(1)$.
This estimate can be directly obtained
as a result of the optimization procedure.

\subsection{Dynamics}

As we will show in the next Section, the UV filtering is very effective in reducing
the number of fields by a factor 3 or more. The improvement may be really impressive for 
heavy quarks, since in this case the small loops account for most of the dynamical effects.

However, as the quark mass decreases, one must include larger loops in the gauge action 
or increase the number of boson fields. Because the multiplicity of larger Wilson
loops increases combinatorically, the best compromise  (see next Section) limits the loop
expansion to small loops and increases the number of fields, which eventually diverges as 
$1/m_q$ \cite{MB1}. In this case the dynamics of the system may be highly non trivial
and the proper choice of the algorithm, or mixture of algorithms, may become very important.

In particular, for light quark masses it is more likely for one of the zeros of the polynomial to get 
very close to the Dirac spectrum boundary, and so
for the corresponding boson field to become almost massless.
The dynamics of those IR boson fields then becomes critical. They represent
the bottleneck of the dynamical evolution of the whole system.
One has then to search for a good algorithm in order to speed up those
slow modes.

Due to the simple form of the bosonic distribution,
\be
P( \phi_k ) \propto {\rm exp} \left( - |(D - z_k) \phi_k |^2 \right) ,
\ee
a global heatbath on the bosonic fields can be simply implemented \cite{borici} 
and turns  out to be very effective.
The new field $\phi_k^{\rm NEW}$ is obtained by applying $(D - z_k)^{-1}$ to
a Gaussian vector $\eta$
\be
\phi_k^{\rm NEW} = (D - z_k)^{-1} \eta .
\ee
In this way $\phi_k^{\rm NEW}$ is completely uncorrelated 
from the old bosonic field.

In practice, since an accurate exact inversion of $(D - z_k)$ is needed,
the global heatbath may become prohibitively expensive,
especially for those fields where $z_k$ is very close to the edge of the Dirac
spectrum, so that $(D - z_k)$ is almost singular.

In order to cure this problem we have adopted, instead of the usual heatbath 
with exact inversion, a quasi-heatbath consisting of an approximate 
inversion plus a Metropolis accept-reject step as proposed in \cite{PhQH}.

The idea is not to perform an exact inversion of $(D - z_k)$, 
but to stop the inversion algorithm (BiCG in our case) early by loosening
the convergence criterion. In order to preserve detailed balance,
the system one solves approximately is
\be
(D - z_k) x = b,
\ee
with right-hand side $b = (D - z_k) \phi_k^{\rm OLD} + \eta$.
The residual is $r = b - (D - z_k) x$.
A candidate bosonic field is then formed as
\be
\phi_k^{\rm NEW} = x - \phi_k^{\rm OLD}.
\ee
It is accepted with probability
\be
min \left( 1,{\rm exp} (2 Re(r^\dagger.(D - z_k) (\phi_k^{\rm NEW} - \phi_k^{\rm OLD}) \right)
\ee
It is easily proven that in this way the bosonic field is sampled
with the correct distribution, for any convergence criterion of the solver. 
Using this simple trick it is possible
to strongly reduce the number of iterations of the inversion
algorithm (by a factor 3 or so) while maintaining an acceptance for
$\phi_k^{\rm NEW}$ close to 1 \cite{PhQH}.
The use of even-odd preconditioning further reduces the computational demand.

This global update of the bosonic fields has then to be combined 
with local update algorithms. We have chosen local overrelaxation
for both gauge and bosonic fields\footnote{Note that no heatbath on gauge 
fields is needed, since ergodicity is already ensured by the global heatbath
on bosonic fields}.

A good tuning of the relative frequencies of the three different algorithms
in the mixture is essential. With the global heatbath some new uncorrelated
information is brought into the statistical system via the boson fields,
which of course needs some time to propagate to the gauge fields too.
While a stochastic choice (with proper weights) among the three algorithms
might seem preferable, we have observed that the use of a deterministic
sequence of algorithms in the trajectory between two successive Metropolis 
steps is much more effective.

\section{Numerical results}

Three representative systems, increasingly demanding in computer resources,
have been studied: medium-heavy quarks in a small lattice, light quarks in a
small lattice, and light quarks in a large lattice. In all three cases the
efficiency of our algorithm at least matched that of HMC. Our simulations
are of length ${\cal O}(10^3) \tau_{int}(\Box)$, sufficient to extract
reasonably accurate (${\cal O}(10\%)$) integrated autocorrelation times 
$\tau_{int}(\Box)$ for the plaquette.

In all cases we simulate 2 flavours of Wilson fermions, and use the non-hermitian
version of the MultiBoson algorithm \cite{MB1} with even-odd preconditioning
and noisy Metropolis correction test. The gauge action is the Wilson plaquette 
action.

We have implemented UV-filtering up to fourth order (i.e. up to
4-link loops), which generates a shift $\Delta\beta$ in the gauge coupling 
$\beta$, but causes no overhead.

The details of the three simulations, denoted (A),(B) and (C), 
are summarized in Table I.
Note that the optimized value of the coefficient $a_4$, and thus $\Delta\beta$,
is much larger
than the hopping parameter expansion value ($1/4$). This is because UV-filtering
does its best to approximate the infinite series given by the hopping 
parameter expansion with a series truncated to 4th order: the best
truncated series is not the truncation of the infinite one.

When performing the Metropolis test between two trajectories the quantity
$W \eta$, where 
$ W \equiv \prod_k^n ({\bf 1} - \kappa M - z_k {\bf 1}) \cdot
({\bf 1} - \kappa M) \cdot e^{\sum_{j=0}^{m-1} a_j M^j} $, has to be computed,
and some care is needed in the evaluation of the exponential
$\xi = e^{\sum_{j=0}^{m-1} a_j M^j} \eta$. Our method is to 
compute it by Taylor expansion, stopping the series when the 
contribution of the first neglected order to $\xi$ is less than
$10^{-14}$ for each site.

For MB simulations (A) and (B), the order of update of the gauge links was
not the usual one. All 8 links attached to a given site $x$, forming a ``star''
pattern, were updated before proceeding to another set of 8 links. This
arrangement allows a re-use of intermediate link-products in the calculation
of the gauge force, thereby reducing the overall amount of computation \cite{star}.
More importantly, it becomes very simple in this scheme to integrate analytically
over the central bosonic fields $\phi_k(x)$, which permits a larger-step update
of the gauge fields. This provides similar advantages to the combined gauge-boson
update of \cite{Beat}, without the overhead.

The HMC simulations used for comparisons incorporate state-of-the-art 
improvements: even-odd preconditioning ((A), (B) and (C)); incomplete convergence of the solver
during the MD trajectory ((A) and (B)) or 
time-extrapolation of the initial guess ((C)); BiCG$\gamma_5$
((A) and (B)) or BiCGStab ((C)) solver; multi-stepsize integration, following 
\cite{Sexton-Weingarten}, in ((A) and (B)). 
Simulation (C) is a SESAM-collaboration production run \cite{SESAM}.
It does not incorporate, however, SSOR preconditioning which they use
with advantage in more recent projects where it reduces the work per
independent configuration by a factor of about two \cite{SSOR}.

\begin{table}[h]
\caption{Summary of the parameters of our 3 simulations. 
The optimized coefficients $a_2$ and $a_4$ and $\Delta\beta$ have been
rounded off to 3 decimals.
Integrated autocorrelation times are measured in applications of the
Wilson Dirac matrix to a vector.}  
\begin{center}
\begin{tabular}{|c|c|c|c|}
Simulation & A & B & C \\
\hline
Volume & $8^4$ & $8^4$ & $16^3 \times 24$ \\
$\beta,\kappa$ & $5.3, 0.158$ & $5.3, 0.165$ & $5.6, 0.156$ \\
$n_{bosons}$ & 7 & 16 & 24 \\
$a_2$ & 4.411 & 6.066 & 4.077 \\
$a_4$ & 1.389 & 4.423 & 8.789 \\
$\Delta\beta=192 \kappa^4 a_4$ & 0.166 & 0.629 & 0.999 \\
$\tau_{int}(\Box)(MB)$ & $\sim 3500$ & $\sim 64000$ & $\sim 27500$ \\
$\tau_{int}(\Box)(HMC)$ & $\sim 14000$ & $\sim 72000$ & $\sim 85000^\dagger$ \\
\end{tabular}
\end{center}
$\dagger$ Obtained from the SESAM-collaboration data on a $16^3 \times 32$
lattice \cite{SESAM}.
\end{table}

\subsection{Medium-heavy quarks, small lattice}
The parameters of the simulation are indicated in Table I, column (A).
Seven auxiliary fields only were required after UV-filtering
(note that there is no need for the number of fields to be even).
Fig.1 shows the location of the associated complex zeros of the approximating
polynomial, together with the boundary of the Dirac spectrum estimated by
diagonalizing the tridiagonal matrix given by the BiCG$\gamma_5$ solver.
Only one zero is devoted to controlling the UV part of the Dirac spectrum, 
while the other 6 control the IR. This is because the shift 
$\Delta\beta \approx 0.166$ in the gauge coupling accounts for most of the UV 
fluctuations already.
Because none of the auxiliary fields has a small mass, as can be judged from the
distance of the zeros to the boundary of the Dirac spectrum Fig.1,
local and global updates of these fields are about equally efficient.
In both cases, the plaquette decorrelates about 4 times faster than with HMC,
as shown in Fig.2.
This good result is inherent to the MB approach: for heavy quarks, it reduces
to pure gauge local updates, whereas HMC remains an infinitesimal-step
algorithm with much slower dynamics \cite{Sharpe}. Therefore, the relevant
question is: at which quark mass, if any, does the MB approach lose its
advantage? This is the reason for our second test.

\begin{figure}[!ht]
\vspace{-0.3cm}
\begin{center}
\epsfig{file=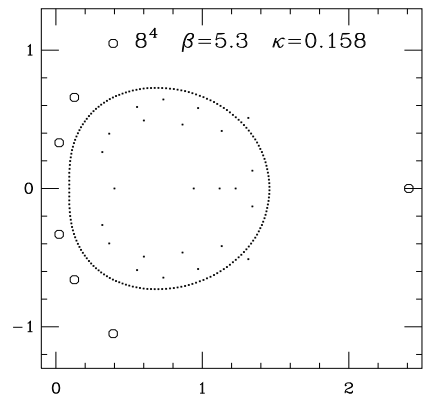,height=5.2cm,width=5.5cm,angle=0}
\vspace{0.2cm}
\caption{Zeros of the UV-filtered polynomial (circles), and estimated boundary
of the Dirac spectrum (crosses), simulation (A).}
\end{center}
\end{figure} 
\vspace{-0.5cm}

\begin{figure}[!h]
\vspace{-0.8cm}
\begin{center}
\epsfig{file=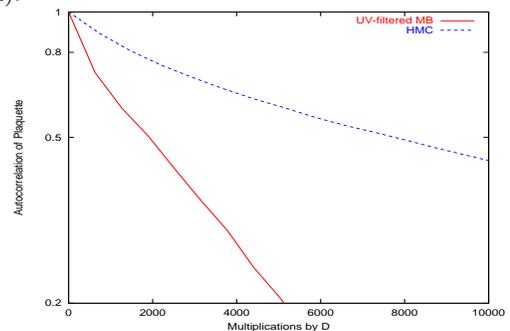,height=4.5cm,width=6.5cm,angle=0}
\caption{Autocorrelation of the plaquette, with
the UV-filtered MB algorithm (solid line) and the HMC algorithm
(dashed line)
at $\beta = 5.3$ and $\kappa = 0.158$ (simulation A).
}
\end{center}
\end{figure} 
\vspace{-0.5cm}

\subsection{Light quarks, small lattice}
The parameters of the simulation are indicated in Table I, column (B).
$\kappa_c$ is approximately 0.1686(3) \cite{Rajan}, so we are simulating light
quarks. As appears clearly in Fig.3, several of the IR zeros are now closer to
the Dirac spectrum boundary. Consequently, the global Quasi-Heatbath provides
a large improvement over local updates, by a factor 5 or more. 
The number of boson fields, $n=16$,
is very much smaller than the number of solver iterations per HMC step ($\sim 53$).
The 2 algorithms appear about equally efficient (see Fig.4). 
Since the work per independent
configuration grows more slowly with the volume for MB than HMC 
($V(Log(V))^2$ vs $V^{5/4}$ \cite{MB1}), this bodes well for realistic
simulations in large volumes.

\begin{figure}[!ht]
\vspace{-0.5cm}
\begin{center}
\epsfig{file=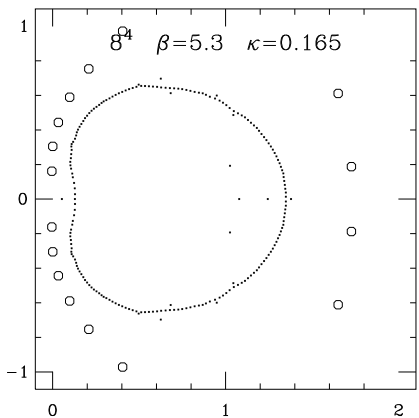,height=5.5cm,width=5.5cm,angle=0}
\vspace{0.2cm}
\caption{Zeros of the UV-filtered polynomial (circles), and estimated boundary
of the Dirac spectrum (crosses), simulation (B).}
\end{center}
\end{figure} 
\vspace{-0.5cm}

\begin{figure}[!ht]
\vspace{-0.8cm}
\begin{center}
\epsfig{file=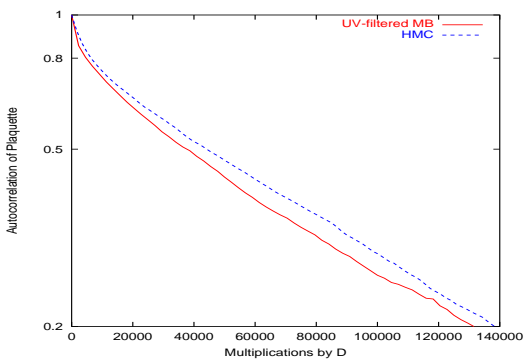,height=5.0cm,width=7cm,angle=0}
\vspace{0.2cm}
\caption{Autocorrelation of the plaquette, with
the UV-filtered MB algorithm (solid line) and the HMC algorithm
(dashed line)
at $\beta = 5.3$ and $\kappa = 0.165$ (simulation B).
}
\end{center}
\end{figure} 
\vspace{-0.5cm}

\subsection{Light quarks, large lattice}

The parameters of the simulation are indicated in Table I, column (C).
This large-scale simulation has been performed on the APE/TOWER machine in Pisa.

The number of boson fields, $n=24$, is again very much smaller than the number
of BiCG iterations per MD step ($\sim 91$). Note that without UV-filtering, these
2 numbers become comparable: we needed $n=80$ fields in this case,
with the zeros of the polynomial evenly spaced on the unit circle centered
at $(1,0)$, to reach similar acceptance ($\sim 74\%$).

Regarding the mixture of algorithms, we have found that a good compromise
is to perform a global heatbath per bosonic field or each symmetrized
trajectory composed of 12 couples of local over-relaxation sweeps for
the bosonic and gauge fields. This choice may be subject to further
optimization.

Of the optimized roots of the UV-filtered polynomial, shown in Fig.5, 
16 are devoted to IR modes and 8 to UV modes of the Dirac operator.

\begin{figure}[!h]
\vspace{0.cm}
\begin{center}
\epsfig{file=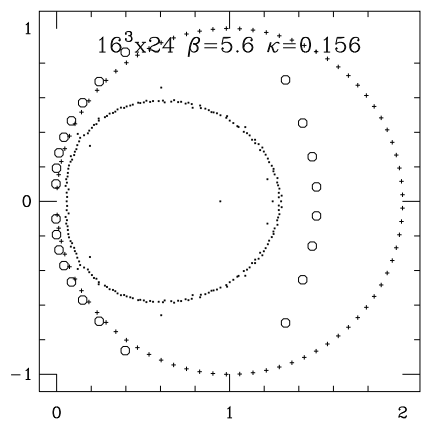,height=5.5cm,width=5.5cm,angle=0}
\vspace{0.2cm}
\caption{Zeros of the UV-filtered polynomial ($n=24$)(circles), of the
non-UV-filtered polynomial ($n=80$)(+), and estimated boundary
of the Dirac spectrum (crosses), simulation (C).}
\end{center}
\end{figure} 
\vspace{0cm}

To illustrate the benefits of UV-filtering and of the bosonic quasi-heatbath,
we show in Fig.6 the evolution of the plaquette, first without UV-filtering
($n=80$ bosonic fields), then with UV-filtering ($n=24$) but without 
quasi-heatbath, and finally with both features. The improvement is clearly
visible in each case. The autocorrelation of the plaquette is compared in
Fig.7 with that obtained by the SESAM collaboration using HMC \cite{SESAM}.
Our MB approach is more efficient by a factor $\sim 3$.

In Fig. 8 we compare the topological charge histories obtained from
our simulation and from a sample of SESAM configurations \cite{HMC2}.
In both cases the same cooling method was used. Neither simulation
is long enough to extract a reliable autocorrelation time for this observable.
Attempts at doing so yield roughly equivalent results for both algorithms,
as the figure already indicates. 
Therefore, even for this global observable,
our MB method seems not worse than HMC. 

\begin{figure}[!ht]
\vspace{-0.5cm}
\begin{center}
\epsfig{file=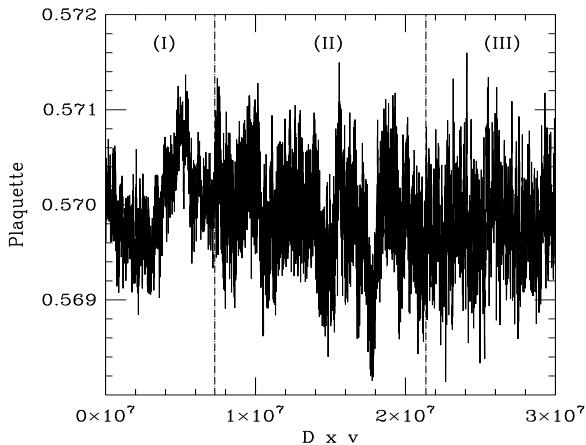,height=5.5cm,width=6.5cm,angle=0}
\vspace{0.4cm}
\caption{Monte Carlo history of the plaquette, with the
non-hermitian MB algorithm (I), then with UV-filtering (II),
finally with UV-filtering and bosonic quasi-heatbath (III) (simulation C).
}
\end{center}
\end{figure} 
\vspace{0cm}

\begin{figure}[!ht]
\vspace{-1.0cm}
\begin{center}
\epsfig{file=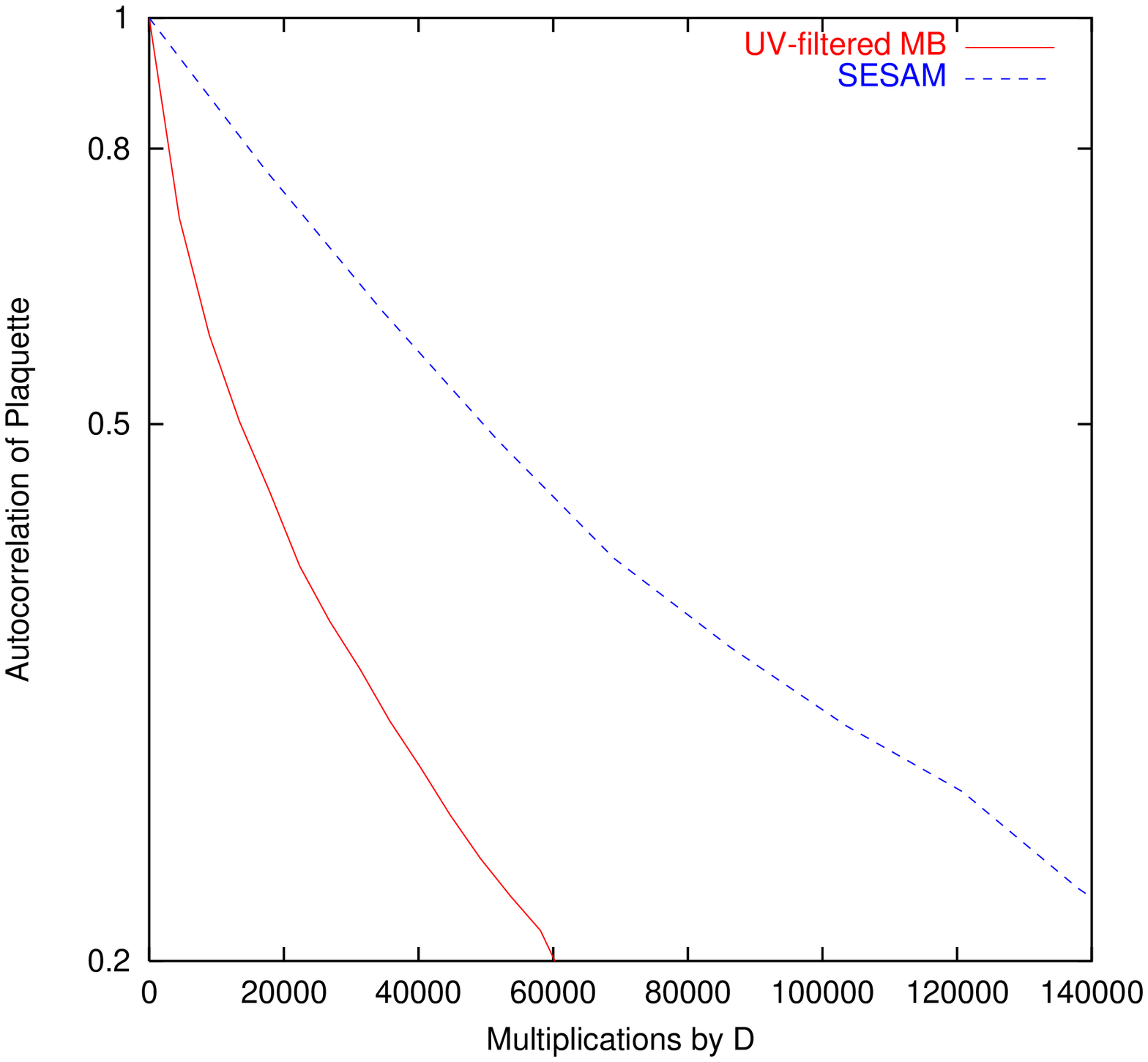,height=5.0cm,width=7cm,angle=0}
\vspace{0.2cm}
\caption{Autocorrelation of the plaquette, with
the UV-filtered MB algorithm (solid line) and the SESAM HMC algorithm
(dashed line)
at $\beta = 5.6$ and $\kappa = 0.156$ (simulation C).
}
\end{center}
\end{figure} 
\vspace{0cm}

\begin{figure}[!ht]
\vspace{-1.0cm}
\begin{center}
\epsfig{file=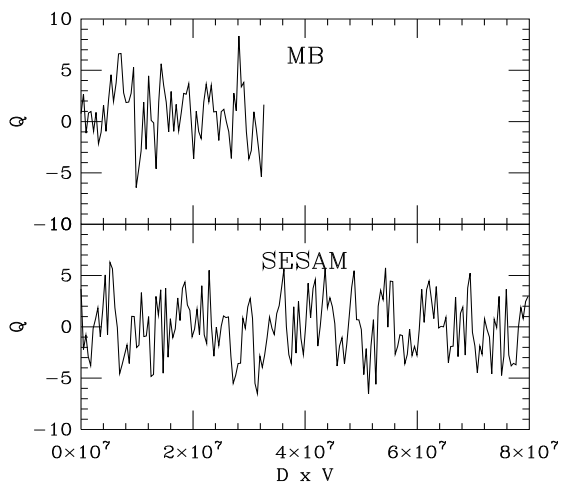,height=6.5cm,width=7.5cm,angle=0}
\vspace{0.2cm}
\caption{Comparison of topological charge histories obtained
with HMC and UV-filtered MB algorithm at $\beta = 5.6$ and $\kappa = 0.156$.
The common scale has been set in terms of equivalent $D\times v$ 
multiplications.}
\end{center}
\end{figure} 
\vspace{0cm}

We considered extending the UV-filtering to order 6. 
This would allow a further reduction of the number of bosonic fields, at the
expense of including in the action 6-link loops coming from $Tr(M^6)$.
The optimization of the UV-filtered polynomial was performed under these premises.
It indicated that the same accuracy obtained with $n=24$ and 4th-order UV-filtering
could be obtained with $n \approx 20$ and 6th-order UV-filtering. 
This relatively small reduction in $n$ did not justify the overhead of including
6-link loops in the update.

\section{Conclusions}

We have presented numerical evidence that the exact, non-hermitian
MB algorithm is a superior alternative to the traditional HMC:
it decorrelates the plaquette more efficiently, and the topological
charge equivalently well.
The key ingredients to achieve this efficiency are UV-filtering and 
global quasi-heatbath of the boson fields. 
The former absorbs the
UV modes of the Dirac operator in the gauge action, and thereby
reduces the required number of bosonic fields by a factor 3 or more, 
thus removing the memory bottleneck of the non-filtered MB.
The latter greatly accelerates, at low computer cost, the dynamics 
of the IR, low-mass boson fields.
The same scheme can be implemented without changes to simulations
with staggered fermions. Similar efficiency gains are expected.
At the same time, the MB polynomial can be tailored to approximate
any number of staggered flavours, for instance $N_f=2$. With a correction step
as used for $N_f=1$ Wilson-quark simulations \cite{Nf1}, this will 
allow for exact, efficient simulation of two staggered flavours, 
which are inaccessible to HMC.

\par\bigskip
    
We thank Adriano Di Giacomo, Klaus Schilling and Thomas Lippert 
for interesting discussions, and the whole SESAM collaboration for sharing
with us their numerical results. 
Ph. de F. thanks Greg Kilcup for access to NERSC computers.
M. D'E. thanks Lele Tripiccione for his valuable help in using
the 512 node APE--QUADRICS in Pisa.
Ph. de F. and M. D'E. thank the Aspen Center for Physics where part of this paper
was written.

\end{document}